\documentclass[%
 preprint,
 showpacs,
 showkeys,
 preprintnumbers,
  amsmath,amssymb,
  aps,
 pra,
  longbibliography,
  floatfix,
 ]{revtex4-1}

\usepackage[breaklinks=true,colorlinks=true,anchorcolor=blue,citecolor=blue,filecolor=blue,menucolor=blue,pagecolor=blue,urlcolor=blue,linkcolor=blue]{hyperref}

\usepackage{graphicx}
\usepackage{epstopdf}
\usepackage{eepic}
\usepackage{xcolor}
\usepackage{braket}
\usepackage{amsmath}
\usepackage{amsthm}

\usepackage{bm}

\RequirePackage{times}
\RequirePackage{mathptm}

\begin{document}


\title{Quantum value indefiniteness}

\author{Karl Svozil}
\email{svozil@tuwien.ac.at}
\homepage{http://tph.tuwien.ac.at/~svozil}
\affiliation{Institut f\"ur Theoretische Physik, Vienna University of Technology,  \\  Wiedner Hauptstra\ss e 8-10/136, A-1040 Vienna, Austria}

\begin{abstract}
The indeterministic outcome of a measurement of an individual quantum is certified by the impossibility of the simultaneous, unique, definite, deterministic pre-existence of all conceivable observables from physical conditions of that quantum alone.
\end{abstract}


\maketitle

\section{Introduction}

One of the most astounding consequences of the assumption of the validity of the quantum formalism in terms
of Hilbert spaces \cite{v-neumann-49} is the apparent impossibility of its classical interpretation.
More precisely, a classical interpretation of a quantum logical structure~\cite{birkhoff-36}
is either identified with a Boolean algebra, or at least with a homomorphic embedding
(structurally preserving all quantum logical relations and operations) into some Boolean algebra~\cite{CalHerSvo}.
Quantum logics are obtained by identifying (unit) vectors
(associated with the one-dimensional subspaces corresponding to the linear spans of the vectors,
and with the corresponding one dimensional projectors) with  elementary yes-no propositions.
The logical {\it and, or,} and {\it not} operations are identified with the set theoretic intersection,
with the linear span of two subspaces, and
with forming the orthogonal subspace, respectively.
Suppose further that orthogonality among subspaces indicates mutual exclusive propositions or
experimental outcomes.

Then, in at least three-dimensional Hilbert (sub)spaces, there does not exist a (classical) truth assignment
on (finite sets of) elementary yes-no propositions which would
\begin{description}
\item[(Rule~1---``countable  additivity:'')]  ascribe truth to
exactly one observable outcome among each set of maximal commeasurable mutually exclusive outcomes,
and falsity to the others,
such that
\item[(Rule~2---``noncontextuality:'')]
for ``overlapping'' link observables belonging to more than one commeasurable set of observables,
henceforth called {\em context,}
the truth value remains the same, independent of the particular commeasured
observables~\cite{specker-60,kochen1,ZirlSchl-65,Alda,Alda2,kamber64,kamber65,peres-91,mermin-93,svozil-tkadlec,cabello-96,cabello:210401}.
\end{description}
Proofs (e.g., \cite{kochen1}) could be finitistic and by contradiction (i.e., {\it via reductio ad absurdum}),
so there should not be any metamathematical issues about their applicability in physics.
Countable additivity (Rule~1) is the basis of
a theorem~\cite{Gleason,pitowsky:218,rich-bridge,r:dvur-93} by Gleason which derives the Born rule
$\langle \textsf{\textbf{A}} \rangle = {\rm Tr}\left({\bm \rho} \textsf{\textbf{A}}\right)$,
where  $\langle \textsf{\textbf{A}} \rangle$ and  ${\bm \rho}$
stand for the expectation value of an observable $\textsf{\textbf{A}}$ and for the quantum state, respectively.

Yet, there are metaphysical issues related to the impossibility of a classical interpretation of the
quantum formalism; in particular the explicit and indispensible use of {\em counterfactuals}
in the argument~\cite{svozil-2006-omni}.
Remarkably, this has been already emphasized in the first announcement of the formal result~\cite{specker-60}.
Counterfactuals are ``observables'' which {\em could} have been measured {\em if}
the experimenter {\em would have} chosen a different, i.e., complementary, measurement setup,
but {\em actually chose another} (complementary) one.
Hence, from the point of view of the quantum formalism,
any proof of the impossibility of a classical interpretation of quantum mechanics
uses complementary observables, which cannot possibly be simultaneously measured.
Pointedly stated, from a strictly operational point of view, due to quantum complementarity~\cite[p.~7]{pauli:58},
the entities occurring in the proofs cannot physically coexist.

So, it may not be totally unjustified to ask
why one should bother about nonoperational quantities and their consequences at all?
There may be two affirmative apologies for the use of counterfactuals:
First, although these observables could not be measured simultaneously,
they are perfectly reasonable physical observables if the experimenter chooses to measure them.
Secondly, through a measurement setup involving two correlated  particles,
two complementary observables can be measured counterfactually~\cite{epr}
on two space-like separated~\cite{wjswz-98}
but entangled~\cite{schrodinger,CambridgeJournals:1737068,CambridgeJournals:2027212} particles.
Because of constraints on the uniqueness of the arguments, this ``indirect measurement'' cannot be
extended to more than two counterfactual observables~\cite{svozil-2006-uniquenessprinciple}.

Quantum ``value (in)definiteness,''  sometimes also termed ``counterfactual (in)definiteness''~\cite{MuBae-90},
refers to the (im)possibility of the simultaneous existence of definite outcomes of conceivable measurements
under certain assumptions [e.g. noncontextuality; see Rule~2 above] --- that is, unperformed measurements can(not)
have definite results~\cite{peres222}.
``(In)determinacy'' often (but not always) refers to the absence (presence) of causal laws
---
in the sense of the principle of sufficient reason stating that every phenomenon has its explanation and cause
---
governing a physical behavior.
Thus ``value (in)definiteness'' relates to a static property, whereas ``(in)determinacy'' is often used for temporal evolutions.
Sometimes, quantum  value indefiniteness is considered as one of the expressions of quantum indeterminacy;
another expression of quantum indeterminacy is, for instance, associated with
the (radioactive) decay of some
excited states~\cite{Kragh-1997AHESradioact,Kragh-2009_RePoss5}.

In what follows we shall review some explicit physical consequences of the impossibility to interpret the quantum formalism classically.
We shall also review consequences for the construction of quantum mechanical devices
capable of generating particular
indeterministic outcomes~\cite{svozil-qct,rarity-94,zeilinger:qct,stefanov-2000,0256-307X-21-10-027,wang:056107,fiorentino:032334,svozil-2009-howto,Kwon:09,10.1038/nature09008},
which have been already discussed in an article~\cite{2008-cal-svo} by Calude and the author.

Any particular maximal set of (mutually exclusive) observables will be called {\em context}~\cite{svozil-2008-ql}.
It constitutes a ``maximal collection of co-measurable observables,'' or, stated differently,
a ``classical mini-universe'' located within the continuity of complementary quantum propositions.
The spectral theorem suggests that a context can be formalized by a single  ``maximal'' self-adjoint operator, such that
there exist ``maximal'' sets of mutually compatible, co-measurable, mutually exclusive orthogonal projectors
which appear in its spectral decomposition
(e.g., \cite[Sec.~II.10, p. 90, English translation p.~173]{v-neumann-49},
\cite[\S~2]{kochen1}, \cite[pp.~227,228]{neumark-54}, and \cite[\S~84]{halmos-vs}).

\section{Contextual interpretation}

In a ``desperate'' attempt to save realism~\cite{stace},
Bell~\cite{bohr-1949,bell-66,hey-red,redhead}
proposed to abandon the noncontextuality assumption Rule~2 that the truth or falsity of an individual outcome of a measurement of some observable
is independent of what other (mutually exclusive) observables are measured ``alongside''
of it.
In Bell's own words~\cite[Sec.~5]{bell-66},
the ``danger'' in the implicit assumption is this\footnote{
Bell cites Bohr's remark~\cite{bohr-1949} about
{\em ``the impossibility of any sharp separation
between the behavior of atomic objects and the interaction with the measuring instruments which serve to define
the conditions under which the phenomena appear.''}}:
\begin{quote}
{\em
``It was tacitly assumed that measurement of an observable must yield the same value independently of
what other measurements may be made simultaneously.
$\ldots$
The result of an observation may reasonably depend
not only on the state of the system  $\ldots$
but also on the complete disposition  of the apparatus.''}
\end{quote}
This ``contextual interpretation'' of quantum mechanics  will be henceforth called {\em contextuality.}

Notice that contextuality
does not suggest that any {\em statistical} property is context dependent;
this would be ruled out by the Born rule, which is context independent.
Instead, the contextual interpretation claims that the {\em individual outcome} --- Bell's ``result of an observation'' ---
depends on the context.
This is somewhat similar to the parameter independence but outcome dependence of correlated quantum events~\cite{shimony2}.

The exact formalization or causes of this type of ``contextual outcome dependence'' remains an open question.
Individual quantum events are generally {\em conventionalized} to happen acausally and indeterministically~\cite{born-26-1,born-26-2};
according to the prevalent quantum canon~\cite{zeil-05_nature_ofQuantum},
{\em ``$\ldots$ for the individual event in quantum physics, not only do we not know the cause, there is no cause.''}
In this belief system, indeterminism can be  trivially certified by the convention of the ``random outcome''
of individual quantum events,
a view which is further
``backed'' by our inability to ``come up'' with a causal model, and by the statistical analysis~\cite{PhysRevA.82.022102} of the assumption
of stochasticity and randomness of strings generated {\em via} the context mismatch between preparation and measurement.
Nevertheless, one should always keep in mind that this kind of indeterminism may be epistemic and not ontic.
Furthermore, due to the ambiguities of a formal definition,
and by reduction to the halting problem~\cite{rogers1,davis,Barwise-handbook-logic,enderton72,odi:89,Boolos-07},
the incomputablity, and even more so randomness, of arbitrary (finite) sequences remains provably unprovable~\cite{calude:02}.

\subsection{Violation of probabilistic bounds}

For the sake of getting a more intuitive understanding of quantum contextuality, a few examples of its consequences will be discussed next.
As any violations of Boole-Bell type elements of physical reality indicate the impossibility of its classical interpretation
by probabilistic constraints~\cite{pitowsky-89a,Pit-94,pitowsky-86,pitowsky},
every violation of Boole-Bell type inequalities can be re-interpreted as (experimental)
``proof of contextuality''~\cite{hasegawa:230401,Bartosik-09,PhysRevLett.103.160405,kirch-09}.
Indeed, as expressed by \cite{cabello:210401},
{\em ``Because of the lack of spacelike separation between one
observer's choice and the other observer's outcome, the
immense majority of the experimental violations of Bell
inequalities does not prove quantum nonlocality, but just
quantum contextuality.''}
Alas, while certainly most (with the exception of, e.g., \cite{wjswz-98})  experimental violations of Bell
inequalities do not prove quantum nonlocality,
these statistical violations are no direct proof of contextuality in general.
Nevertheless, they may indicate counterfactual indefiniteness~\cite{MuBae-90}.

Note that in a geometric framework~\cite{froissart-81,cirelson:80,cirelson,pitowsky-89a,Pit-94,pitowsky-86,pitowsky,2000-poly},
Boole-Bell type inequalities are just the {\em facet inequalities} of a classical probability (correlation) polytope
obtained by (i) forming all probabilities and joint probabilities of independent events,
(ii) taking all two-valued measures (interpretable as truth assignments) associated with this structure,
(iii) for each of the probabilities and joint probabilities forming a vector whose components are the (encoded truth) values
(either ``$0$'' or ``$1$'') of the two-valued measures
(hence, the dimensionality of the problem is equal to the number of entries corresponding to  probabilities and joint probabilities);
every such vector is a vertex of the {\em correlation polytope},
(iv) applying the
Minkoswki-Weyl representation theorem (e.g., \cite[p.29]{ziegler}),  stating that
every convex polytope has a dual (equivalent) description  as the intersection of a finite number of half-spaces.
Such facets are given by linear inequalities,
which are obtained
from the set of vertices
by solving the (computationally hard \cite{pit:90})
{\em hull problem}.
The inequalities coincide with Boole's {\em ``conditions of possible experience,''} and with Bell type inequalities.

Any  ``proof'' of contextuality based on Boole-Bell type inequalities necessarily involves the {\em statistical} behavior of
many counterfactual quantities  contained in  Boole-Bell type inequalities.
These quantities cannot be obtained simultaneously,
but merely one after another in different experimental configuration runs
involving ``lots of particles.''
Due to the statistical nature of the argument and its implicit
improvable assumption that contextuality --- that is, the abandonment of Rule~2 --- is the only possible cause for
the violations of the classical probabilistic bounds, these ``proofs'' lack the {\em sufficiency} of the formal argument.

\subsection{Tables of counterfactual ``outcomes''}

Previously,  tables of hypothetical and counterfactual experimental outcomes
have been used to argue against the noncontextual classical interpretation
of the quantum probabilities~\cite{peres222,MuBae-90,krenn1}.
In what follows tables of contextual outcomes violating Rule~2 will be enumerated which could be compatible
with quantum probabilities.
These tables may serve as a demonstration of the kind of behavior which is required by (hypothetical and counterfactual)
individual events capable of rendering the desired violations of Boole-Bell type violations of bounds on classical probabilities.

Let ``$\textsf{\textbf{X}} \{  \textsf{\textbf{Y}} \}  $'' stand for ``observable $\textsf{\textbf{X}}$
measured alongside observable (or context) $\textsf{\textbf{Y}}$.''
Consider the hypothetical counterfactual outcomes enumerated in Table~\ref{2010-pc09-t1}
for simultaneous quantum observables associated with the  Clauser-Horne-Shimony-Holt inequality
\begin{equation}
\vert
\textsf{\textbf{A}}_1 \{  \textsf{\textbf{B}}_1 \} \textsf{\textbf{B}}_1 \{  \textsf{\textbf{A}}_1 \}  +
\textsf{\textbf{A}}_1 \{  \textsf{\textbf{B}}_2 \} \textsf{\textbf{B}}_2 \{  \textsf{\textbf{A}}_1 \}  +
\textsf{\textbf{A}}_2 \{  \textsf{\textbf{B}}_1 \} \textsf{\textbf{B}}_1 \{  \textsf{\textbf{A}}_2 \}  -
\textsf{\textbf{A}}_2 \{  \textsf{\textbf{B}}_2 \} \textsf{\textbf{B}}_2 \{  \textsf{\textbf{A}}_2 \}
\vert \le 2.
\end{equation}
They are contextual, as for some cases $\textsf{\textbf{X}} \{  \textsf{\textbf{Y}}_1 \}  \neq  \textsf{\textbf{X}} \{  \textsf{\textbf{Y}}_2 \}  $, as indicated in the enumeration.
(Note that noncontextuality would imply the independence of $\textsf{\textbf{X}}$ from $\textsf{\textbf{Y}}$; i.e.,
$\textsf{\textbf{X}} \{  \textsf{\textbf{Y}}_1 \}  =  \textsf{\textbf{X}} \{  \textsf{\textbf{Y}}_2 \}  =\textsf{\textbf{X}}$.)
\begin{table}
\begin{center}
\begin{tabular}{l ccccccccccccccccccccccccccccccccccc}
\hline\hline
$\textsf{\textbf{A}}_1 \{  \textsf{\textbf{B}}_1 \}  $ & & $+$       &  $\cdot$   &   $\cdot$ &    $\cdot$ &  $\cdot$   &   $\cdot$    & $\cdot$   &  $\cdot$   &   $\cdot$ &    $\cdot$ &  $+$ &   $\cdot$ &     & $\cdots$  \\
$\textsf{\textbf{A}}_1 \{  \textsf{\textbf{B}}_2 \}  $ & & $-$       &  $\cdot$   &   $\cdot$ &    $\cdot$ &  $\cdot$   &   $\cdot$    & $\cdot$   &  $\cdot$   &   $\cdot$ &    $\cdot$ &  $-$ &   $\cdot$ &     & $\cdots$  \\
$\textsf{\textbf{A}}_2 \{  \textsf{\textbf{B}}_1 \}  $ & & $\cdot$   &  $-$       &   $\cdot$ &    $\cdot$ &  $\cdot$   &   $\cdot$    & $+$       &  $\cdot$   &   $\cdot$ &    $\cdot$ &  $-$ &   $\cdot$ &     & $\cdots$  \\
$\textsf{\textbf{A}}_2 \{  \textsf{\textbf{B}}_2 \}  $ & & $\cdot$   &  $+$       &   $\cdot$ &    $\cdot$ &  $\cdot$   &   $\cdot$    & $-$       &  $\cdot$   &   $\cdot$ &    $\cdot$ &  $+$ &   $\cdot$ &     & $\cdots$  \\
$\textsf{\textbf{B}}_1 \{  \textsf{\textbf{A}}_1 \}  $ & & $\cdot$   &  $\cdot$   &   $\cdot$ &    $\cdot$ &  $-$       &   $\cdot$    & $\cdot$   &  $\cdot$   &   $\cdot$ &    $\cdot$ &  $-$ &   $+$ &         & $\cdots$  \\
$\textsf{\textbf{B}}_1 \{  \textsf{\textbf{A}}_2 \}  $ & & $\cdot$   &  $\cdot$   &   $\cdot$ &    $\cdot$ &  $+$       &   $\cdot$    & $\cdot$   &  $\cdot$   &   $\cdot$ &    $\cdot$ &  $+$ &   $-$ &         & $\cdots$  \\
$\textsf{\textbf{B}}_2 \{  \textsf{\textbf{A}}_1 \}  $ & & $\cdot$   &  $\cdot$   &   $\cdot$ &    $\cdot$ &  $+$       &   $\cdot$    & $\cdot$   &  $-$       &   $\cdot$ &    $\cdot$ &  $-$ &   $\cdot$ &     & $\cdots$  \\
$\textsf{\textbf{B}}_2 \{  \textsf{\textbf{A}}_2 \}  $ & & $\cdot$   &  $\cdot$   &   $\cdot$ &    $\cdot$ &  $-$       &   $\cdot$    & $\cdot$   &  $+$       &   $\cdot$ &    $\cdot$ &  $+$ &   $\cdot$ &     & $\cdots$  \\
\hline\hline
\end{tabular}
\end{center}
\caption{Hypothetical counterfactual contextual outcomes of an experiment capable of violating the Boole-Bell type inequalities
involving binary outcomes (denoted by ``$-$, $+$'') of two observables (subscripts ``1, 2'') on two particles (denoted by ``$\textsf{\textbf{A}}$, $\textsf{\textbf{B}}$'').
The expression ``$\textsf{\textbf{X}} \{  \textsf{\textbf{Y}} \}  $''
stands for ``observable $\textsf{\textbf{X}}$ measured alongside observable $\textsf{\textbf{Y}}$.''
Time progresses from left to right;
rows contain the individual conceivable, potential measurement values of the eight observables
$\textsf{\textbf{A}}_1 \{  \textsf{\textbf{B}}_1 \}  $,
$\textsf{\textbf{A}}_1 \{  \textsf{\textbf{B}}_2 \}  $,
$\textsf{\textbf{A}}_2 \{  \textsf{\textbf{B}}_1 \}  $,
$\textsf{\textbf{A}}_2 \{  \textsf{\textbf{B}}_2 \}  $,
$\textsf{\textbf{B}}_1 \{  \textsf{\textbf{A}}_1 \}  $,
$\textsf{\textbf{B}}_1 \{  \textsf{\textbf{A}}_2 \}  $,
$\textsf{\textbf{B}}_2 \{  \textsf{\textbf{A}}_1 \}  $,  and
$\textsf{\textbf{B}}_2 \{  \textsf{\textbf{A}}_2 \}  $
which ``simultaneously co-exist.''
Dots indicate any value in $\{-,+\}$.
\label{2010-pc09-t1} }
\end{table}

The difference between ``truth tables''
associated with configurations
for the statistical arguments against value indefiniteness involving Boole-Bell type inequalities
on the one hand,
and for direct proofs (e.g. by the Kochen-Specker theorem)
on the other hand,
is that the former tables need not always contain contextual assignments
---
although it can be expected that the violations of noncontextuality should
increase with increasing deviations from the classical Boole-Bell bounds on joint probabilities\footnote{
Note that for stronger-than-quantum correlations~\cite{pop-rohr,svozil-krenn} rendering a maximal
violation of the Clauser-Horne-Shimony-Holt inequality by
$
\textsf{\textbf{A}}_1 \{  \textsf{\textbf{B}}_1 \} \textsf{\textbf{B}}_1 \{  \textsf{\textbf{A}}_1 \}  +
\textsf{\textbf{A}}_1 \{  \textsf{\textbf{B}}_2 \} \textsf{\textbf{B}}_2 \{  \textsf{\textbf{A}}_1 \}  +
\textsf{\textbf{A}}_2 \{  \textsf{\textbf{B}}_1 \} \textsf{\textbf{B}}_1 \{  \textsf{\textbf{A}}_2 \}  -
\textsf{\textbf{A}}_2 \{  \textsf{\textbf{B}}_2 \} \textsf{\textbf{B}}_2 \{  \textsf{\textbf{A}}_2 \}  = \pm 4
$,
if
$\textsf{\textbf{A}}_1 \{  \textsf{\textbf{B}}_2 \} = \textsf{\textbf{A}}_2 \{  \textsf{\textbf{B}}_2 \}$,
then
$\textsf{\textbf{B}}_2 \{  \textsf{\textbf{A}}_1 \} = -\textsf{\textbf{B}}_2 \{  \textsf{\textbf{A}}_2 \}$,
and
if
$\textsf{\textbf{B}}_1 \{  \textsf{\textbf{A}}_2 \} = \textsf{\textbf{B}}_2 \{  \textsf{\textbf{A}}_2 \}$,
then
$\textsf{\textbf{A}}_2 \{  \textsf{\textbf{B}}_1 \} = -\textsf{\textbf{A}}_2 \{  \textsf{\textbf{B}}_2 \}$.}
---
whereas the latter tables require some violation(s) of noncontextuality at every single column.
For example, in the compact 18-vector configuration allowing a Kochen-Specker proof
introduced in~\cite{cabello-96,cabello-99} and depicted in Fig.~\ref{2007-miracles-ksc},
one is forced to violate the noncontextuality assumption Rule~2 for at least one link observable.
This can be readily demonstrated by considering all 36 entries per column in Table~\ref{2010-pc09-t2},
Whether one violation of the noncontextuality Rule~2 is enough for consistency (i.e., the necessary extent of the violation of contextuality)
with the quantum probabilities remains unknown.
\begin{figure}
\begin{center}
\unitlength .6mm 
\allinethickness{2pt} 
\ifx\plotpoint\undefined\newsavebox{\plotpoint}\fi 
\begin{picture}(134.09,125.99)(0,0)

\multiput(86.39,101.96)(.119617225,-.208133971){209}{{\color{green}\line(0,-1){0.208133971}}}
\multiput(86.39,14.96)(.119617225,.208133971){209}{{\color{red}\line(0,1){0.208133971}}}
\multiput(36.47,101.96)(-.119617225,-.208133971){209}{{\color{yellow}\line(0,-1){0.208133971}}}
\multiput(36.47,14.96)(-.119617225,.208133971){209}{{\color{magenta}\line(0,1){0.208133971}}}
\color{blue}\put(86.39,15.21){\color{blue}\line(-1,0){50}}
\put(86.39,101.71){\color{violet}\line(-1,0){50}}
\put(36.34,15.16){\color{magenta}\circle{6}}
\put(36.34,15.16){\color{blue}\circle{4}}
\put(52.99,15.16){\color{blue}\circle{4}}
\put(52.99,15.16){\color{cyan}\circle{6}}
\put(69.68,15.16){\color{blue}\circle{4}}
\put(69.68,15.16){\color{orange}\circle{6}}
\put(86.28,15.16){\color{blue}\circle{4}}
\put(86.28,15.16){\color{red}\circle{6}}
\put(93.53,27.71){\color{red}\circle{4}}
\put(93.53,27.71){\color{orange}\circle{6}}
\put(102.37,43.44){\color{red}\circle{4}}
\put(102.37,43.44){\color{olive}\circle{6}}
\put(111.21,58.45){\color{red}\circle{4}}
\color{green}\put(111.21,58.45){\circle{6}}
\put(102.37,73.47){\color{green}\circle{4}}
\put(102.37,73.47){\color{olive}\circle{6}}
\put(93.53,89.21){\color{green}\circle{4}}
\put(93.53,89.21){\color{cyan}\circle{6}}
\put(86.28,101.76){\color{green}\circle{4}}
\put(86.28,101.76){\color{violet}\circle{6}}
\put(69.68,101.76){\color{violet}\circle{4}}
\put(69.68,101.76){\color{cyan}\circle{6}}
\put(52.99,101.76){\color{violet}\circle{4}}
\put(52.99,101.76){\color{orange}\circle{6}}
\put(36.34,101.76){\color{violet}\circle{4}}
\put(36.34,101.76){\color{yellow}\circle{6}}
\put(29.24,89.21){\color{yellow}\circle{4}}
\put(29.24,89.21){\color{orange}\circle{6}}
\put(20.4,73.47){\color{yellow}\circle{4}}
\put(20.4,73.47){\color{olive}\circle{6}}
\put(11.56,58.45){\color{yellow}\circle{4}}
\put(11.56,58.45){\color{magenta}\circle{6}}

\put(20.4,43.44){\color{magenta}\circle{4}}
\put(20.4,43.44){\color{olive}\circle{6}}
\put(29.24,27.71){\color{magenta}\circle{4}}
\put(29.24,27.71){\color{cyan}\circle{6}}

\color{cyan}
\qbezier(29.2,27.73)(23.55,-5.86)(52.99,15.24)
\qbezier(29.2,27.88)(36.93,75)(69.63,101.91)
\qbezier(52.69,15.24)(87.47,40.96)(93.72,89.27)
\qbezier(93.72,89.27)(98.4,125.99)(69.49,102.06)
\color{orange}
\qbezier(93.57,27.73)(99.22,-5.86)(69.78,15.24)
\qbezier(93.57,27.88)(85.84,75)(53.13,101.91)
\qbezier(70.08,15.24)(35.3,40.96)(29.05,89.27)
\qbezier(29.05,89.27)(24.37,125.99)(53.28,102.06)
\color{olive}
\qbezier(20.15,73.72)(-11.67,58.52)(20.15,43.31)
\qbezier(20.33,73.72)(61.34,93.16)(102.36,73.72)
\qbezier(102.36,73.72)(134.09,58.52)(102.53,43.31)
\qbezier(102.53,43.31)(60.99,23.43)(20.15,43.49)
{\color{black}
\put(30.41,114.02){\makebox(0,0)[cc]{$\textsf{\textbf{M}}$}}
\put(30.41,2.65){\makebox(0,0)[cc]  {$\textsf{\textbf{A}}$}}
\put(52.68,114.38){\makebox(0,0)[cc]{$\textsf{\textbf{L}}$}}
\put(52.68,2.3){\makebox(0,0)[cc]   {$\textsf{\textbf{B}}$}}
\put(91.93,114.2){\makebox(0,0)[cc] {$\textsf{\textbf{J}}$}}
\put(91.93,2.48){\makebox(0,0)[cc]  {$\textsf{\textbf{D}}$}}
\put(69.65,114.38){\makebox(0,0)[cc]{$\textsf{\textbf{K}}$}}
\put(73.65,2.3){\makebox(0,0)[cc]   {$\textsf{\textbf{C}}$}}
\put(103.24,94.22){\makebox(0,0)[cc]{$\textsf{\textbf{I}}$}}
\put(17.45,94.22){\makebox(0,0)[cc] {$\textsf{\textbf{N}}$}}
\put(106.24,22.45){\makebox(0,0)[cc]{$\textsf{\textbf{E}}$}}
\put(17.45,22.45){\makebox(0,0)[cc] {$\textsf{\textbf{R}}$}}
\put(115.13,77.96){\makebox(0,0)[cc]{$\textsf{\textbf{H}}$}}
\put(8.55,77.96){\makebox(0,0)[cc]  {$\textsf{\textbf{O}}$}}
\put(115.13,38.72){\makebox(0,0)[cc]{$\textsf{\textbf{F}}$}}
\put(10.55,38.72){\makebox(0,0)[cc] {$\textsf{\textbf{Q}}$}}
\put(120.92,57.98){\makebox(0,0)[l] {$\textsf{\textbf{G}}$}}
\put(1.77,57.98){\makebox(0,0)[rc]  {$\textsf{\textbf{P}}$}}
}
\put(61.341,9.192){\color{blue}\makebox(0,0)[cc]    {$\textsf{\textbf{a}}$}}
\put(102.883,35.355){\color{red}\makebox(0,0)[cc]   {$\textsf{\textbf{b}}$}}
\put(102.53,84.322){\color{green}\makebox(0,0)[cc]  {$\textsf{\textbf{c}}$}}
\put(60.457,108.01){\color{violet}\makebox(0,0)[cc] {$\textsf{\textbf{d}}$}}
\put(18.031,84.145){\color{yellow}\makebox(0,0)[cc] {$\textsf{\textbf{e}}$}}
\put(18.561,33.057){\color{magenta}\makebox(0,0)[cc]{$\textsf{\textbf{f}}$}}
\put(61.341,39.774){\color{olive}\makebox(0,0)[cc]  {$\textsf{\textbf{g}}$}}
\put(72.124,67.882){\color{orange}\makebox(0,0)[cc] {$\textsf{\textbf{h}}$}}
\put(48.79,67.705){\color{cyan}\makebox(0,0)[cc]    {$\textsf{\textbf{i}}$}}
\end{picture}
\end{center}
\caption{(Color online) Greechie diagram of a finite subset of the continuum of blocks or contexts embeddable in
four-dimensional real Hilbert space without a two-valued probability measure~\protect\cite{cabello-96,cabello-99}  .
The proof of the Kochen-Specker theorem  uses  nine tightly interconnected contexts
$\color{blue}    \textsf{\textbf{a}}=\{\textsf{\textbf{A}},\textsf{\textbf{B}},\textsf{\textbf{C}},\textsf{\textbf{D}}\}$,
$\color{red}     \textsf{\textbf{b}}=\{\textsf{\textbf{D}},\textsf{\textbf{E}},\textsf{\textbf{F}},\textsf{\textbf{G}}\}$,
$\color{green}   \textsf{\textbf{c}}=\{\textsf{\textbf{G}},\textsf{\textbf{H}},\textsf{\textbf{I}},\textsf{\textbf{J}}\}$,
$\color{violet}  \textsf{\textbf{d}}=\{\textsf{\textbf{J}},\textsf{\textbf{K}},\textsf{\textbf{L}},\textsf{\textbf{M}}\}$,
$\color{yellow}  \textsf{\textbf{e}}=\{\textsf{\textbf{M}},\textsf{\textbf{N}},\textsf{\textbf{O}},\textsf{\textbf{P}}\}$,
$\color{magenta} \textsf{\textbf{f}}=\{\textsf{\textbf{P}},\textsf{\textbf{Q}},\textsf{\textbf{R}},\textsf{\textbf{A}}\}$,
$\color{orange}  \textsf{\textbf{g}}=\{\textsf{\textbf{B}},\textsf{\textbf{I}},\textsf{\textbf{K}},\textsf{\textbf{R}}\}$,
$\color{olive}   \textsf{\textbf{h}}=\{\textsf{\textbf{C}},\textsf{\textbf{E}},\textsf{\textbf{L}},\textsf{\textbf{N}}\}$,
$\color{cyan}    \textsf{\textbf{i}}=\{\textsf{\textbf{F}},\textsf{\textbf{H}},\textsf{\textbf{O}},\textsf{\textbf{Q}}\}$
consisting of the 18 projectors associated with the one dimensional subspaces spanned by
$ \textsf{\textbf{A}}=(0,0,1,-1)    $,
$ \textsf{\textbf{B}}=(1,-1,0,0)    $,
$ \textsf{\textbf{C}}=(1,1,-1,-1)   $,
$ \textsf{\textbf{D}}=(1,1,1,1)     $,
$ \textsf{\textbf{E}}=(1,-1,1,-1)  $,
$ \textsf{\textbf{F}}=(1,0,-1,0)   $,
$ \textsf{\textbf{G}}=(0,1,0,-1)   $,
$ \textsf{\textbf{H}}=(1,0,1,0)    $,
$ \textsf{\textbf{I}}=(1,1,-1,1)   $,
$ \textsf{\textbf{J}}=(-1,1,1,1)    $,
$ \textsf{\textbf{K}}=(1,1,1,-1)    $,
$ \textsf{\textbf{L}}=(1,0,0,1)     $,
$ \textsf{\textbf{M}}=(0,1,-1,0)    $,
$ \textsf{\textbf{N}}=(0,1,1,0)    $,
$ \textsf{\textbf{O}}=(0,0,0,1)    $,
$ \textsf{\textbf{P}}=(1,0,0,0)    $,
$ \textsf{\textbf{Q}}=(0,1,0,0)    $,
$ \textsf{\textbf{R}}=(0,0,1,1)    $.
Greechie diagram representing atoms by points, and  contexts by maximal smooth, unbroken curves.
Every observable proposition occurs in exactly two contexts.
Thus, in an enumeration of the four observable propositions of each of the nine contexts,
there appears to be an {\em even} number of true propositions.
Yet, as there is an odd number of contexts,
there should be an {\em odd} number (actually nine) of true propositions.   \label{2007-miracles-ksc} }
\end{figure}

\begin{table}
\begin{center}
\begin{tabular}{l ccccccccccccccccccccccccccccccccccc}
\hline\hline
\color{blue}$\textsf{\textbf{A}} \{   \textsf{\textbf{a}}  \} $   & & $1$   &  $0$   &   $0$ &    $0$ &  $0$   &   $1$    & $0$   &  $1$   &  $0$ &    $0$ &  $0$   &   $1$  &   $1$         & $\cdots$  \\
\color{magenta} $\textsf{\textbf{A}} \{  \textsf{\textbf{f}} \}  $ & & $0$   &  $1$   &   $0$ &    $0$ &  $0$   &   $1$    & $0$   &  $1$   &   $0$ &    $0$ &  $0$ &   $1$   &   $1$         & $\cdots$  \\
\color{blue}$\textsf{\textbf{B}} \{ \textsf{\textbf{a}}  \}  $    & & $0$   &  $0$   &   $1$ &    $0$ &  $0$   &   $0$    & $1$   &  $0$   &  $1$ &    $0$ &  $0$   &   $0$  &   $0$         & $\cdots$  \\
\color{cyan} $\textsf{\textbf{B}} \{  \textsf{\textbf{i}} \}  $ & &$\cdot$&  $0$       &   $\cdot$ &    $0$     &  $0    $   &   $\cdot$    & $\cdot$       &  $0$       &   $\cdot$ &    $\cdot$ &  $0$     &   $0    $ & $\cdot$& $\cdots$  \\
\color{blue}$\textsf{\textbf{C}} \{ \textsf{\textbf{a}} \} $    & & $0$   &  $1$   &   $0$ &    $0$ &  $1$   &   $0$    & $0$   &  $0$   &  $0$ &    $0$ &  $1$   &   $0$  &   $0$         & $\cdots$  \\
\color{olive} $\textsf{\textbf{C}} \{  \textsf{\textbf{h}} \}  $ & & $\cdot$   &  $\cdot$   &   $\cdot$ &    $\cdot$ &  $\cdot$       &   $\cdot$    & $\cdot$   &  $\cdot$   &   $\cdot$ &    $\cdot$ &  $\cdot$ &   $\cdot$ &  $\cdot$& $\cdots$  \\
\color{blue}$\textsf{\textbf{D}} \{ \textsf{\textbf{a}} \} $    & & $0$   &  $0$   &   $0$ &    $1$ &  $0$   &   $0$    & $0$   &  $0$   &  $0$ &    $1$ &  $0$   &   $0$  &   $0$         & $\cdots$ \\
\color{red} $\textsf{\textbf{D}} \{  \textsf{\textbf{b}} \}  $ & & $\cdot$   &  $\cdot$   &   $\cdot$ &    $\cdot$ &  $\cdot$       &   $\cdot$    & $\cdot$   &  $\cdot$       &   $\cdot$ &    $\cdot$ &  $\cdot$ &   $\cdot$ & $\cdot$ & $\cdots$  \\
$\cdots$  & & $\cdot$   &  $\cdot$   &   $\cdot$ &    $\cdot$ &  $\cdot$       &   $\cdot$    & $\cdot$   &  $\cdot$       &   $\cdot$ &    $\cdot$ &  $\cdot$ &   $\cdot$ &   $\cdot$   & $\cdots$  \\
\color{yellow} $\textsf{\textbf{P}} \{  \textsf{\textbf{e}} \}  $ & & $\cdot$       &  $\cdot$   &   $\cdot$ &    $\cdot$ &  $\cdot$   &   $\cdot$    & $\cdot$   &  $\cdot$   &   $\cdot$ &    $\cdot$ &  $\cdot$ &   $\cdot$ &  $\cdot$& $\cdots$  \\
\color{magenta} $\textsf{\textbf{P}} \{  \textsf{\textbf{f}} \}  $ & & $0$   &  $0$   &   $0$ &    $1$ &  $0$   &   $0$    & $1$   &  $0$   &   $0$ &    $1$ &  $0$ &   $0$   &   $0$         & $\cdots$  \\
\color{orange} $\textsf{\textbf{Q}} \{  \textsf{\textbf{g}} \}  $ & & $\cdot$   &  $\cdot$       &   $\cdot$ &    $\cdot$ &  $\cdot$   &   $\cdot$    & $\cdot$       &  $\cdot$   &   $\cdot$ &    $\cdot$ &  $\cdot$ &   $\cdot$ & $\cdot$& $\cdots$  \\
\color{magenta} $\textsf{\textbf{Q}} \{  \textsf{\textbf{f}} \}  $ & & $1$   &  $0$   &   $0$ &    $0$ &  $0$   &   $0$    & $0$   &  $0$   &   $1$ &    $0$ &  $0$ &   $0$   &   $0$         & $\cdots$  \\
\color{cyan} $\textsf{\textbf{R}} \{  \textsf{\textbf{i}} \}  $ & & $0$   &  $\cdot$   &   $0$     &    $\cdot$ &  $\cdot$   &   $0$        & $0$           &  $\cdot$   &   $0$     &    $0$     &  $\cdot$ &   $\cdot$ &  $0$&    $\cdots$  \\
\color{magenta} $\textsf{\textbf{R}} \{  \textsf{\textbf{f}} \}  $ & & $0$   &  $0$   &   $1$ &    $0$ &  $1$   &   $0$    & $0$   &  $0$   &   $0$ &    $0$ &  $1$ &   $0$   &   $0$         & $\cdots$  \\
\hline\hline
\end{tabular}
\end{center}
\caption{(Color online) Hypothetical counterfactual contextual outcomes of  experiments associated
with a compact proof of the Kochen-Specker theorem~\protect\cite{cabello-96,cabello-99}
involving binary outcomes  ``$0$'' or $1$'' of 18 observables,  adding up to one within each of the nine contexts
denoted by ``$\textsf{\textbf{a}}$, $\ldots$, $\textsf{\textbf{i}}$''.
The expression ``$\textsf{\textbf{X}} \{  \textsf{\textbf{y}} \}  $''
stands for ``observable $\textsf{\textbf{X}}$ measured alongside the context $\textsf{\textbf{y}}$.''
Time progresses from left to right;
rows contain the individual conceivable, potential measurement values of the  observables
$\textsf{\textbf{A}} \{  \textsf{\textbf{a}} \}  , \ldots  ,\textsf{\textbf{R}} \{  \textsf{\textbf{i}} \}  $
which ``simultaneously co-exist.''
Dots indicate any value in $\{0,1\}$
subject to at least one violation of the noncontextuality assumption,
that is, $\textsf{\textbf{X}}(\textsf{\textbf{y}}) \neq \textsf{\textbf{X}}(\textsf{\textbf{y}}')$
for $\textsf{\textbf{y}} \neq \textsf{\textbf{y}}'$.
\label{2010-pc09-t2} }
\end{table}

If such signatures of contextuality exist cannot be decided experimentally, as direct observations
are operationally blocked by quantum complementarity. Thus this type of contextuality remains metaphysical.

\subsection{Indirect simultaneous tests}

\begin{figure}
\begin{center}
\begin{tabular}{ccccc}
\unitlength 0.3mm 
\allinethickness{2pt}
\ifx\plotpoint\undefined\newsavebox{\plotpoint}\fi 
\begin{picture}(132.5,122)(0,0)
\put(20,20){\color{green}\line(1,0){73}}
\multiput(20,20)(.03372164316,.05518087063){1080}{\color{red}\line(0,1){0.05518087063}}
\put(20,20){\color{red}\circle{5.5}}
\put(20,20){\color{red}\circle{1.5}}
\put(20,20){\color{green}\circle{9}}
\put(56.25,20){\color{green}\circle{5.5}}
\put(56.25,20){\color{green}\circle{1.5}}
\put(92.5,20){\color{green}\circle{5.5}}
\put(92.5,20){\color{green}\circle{1.5}}
\put(56.25,79.75){\color{red}\circle{5.5}}
\put(56.25,79.75){\color{red}\circle{1.5}}
\put(38.75,51.25){\color{red}\circle{5.5}}
\put(38.75,51.25){\color{red}\circle{1.5}}
\put(15,5){\makebox(0,0)[cc]{$\textsf{\textbf{A}}$}}
\put(56.25,5){\makebox(0,0)[cc]{{\color{green}$\textsf{\textbf{B}}$}}}
\put(92.5,5){\makebox(0,0)[cc]{{\color{green}$\textsf{\textbf{C}}$}}}
\put(28,52.25){\makebox(0,0)[rc]{{\color{red}$\textsf{\textbf{D}}$}}}
\put(42.5,80.25){\makebox(0,0)[rc]{{\color{red}$\textsf{\textbf{E}}$}}}
\end{picture}
&
\unitlength 0.2mm 
\allinethickness{2pt}
\ifx\plotpoint\undefined\newsavebox{\plotpoint}\fi 
\begin{picture}(132.5,122)(0,0)
\put(20,20){\color{green}\line(1,0){110}}
\multiput(20,20)(.03372164316,.05518087063){1631}{\color{blue}\line(0,1){0.05518087063}}
\multiput(75,110)(.03372164316,-.05518087063){1631}{\color{red}\line(0,-1){0.05518087063}}
\put(20,20){\color{blue}\circle{5.5}}
\put(20,20){\color{blue}\circle{1.5}}
\put(20,20){\color{green}\circle{9}}
\put(56.25,20){\color{green}\circle{5.5}}
\put(56.25,20){\color{green}\circle{1.5}}
\put(92.5,20){\color{green}\circle{5.5}}
\put(92.5,20){\color{green}\circle{1.5}}
\put(129.75,20){\color{green}\circle{5.5}}
\put(129.75,20){\color{green}\circle{1.5}}
\put(129.75,20){\color{red}\circle{9}}
\put(56.25,79.75){\color{blue}\circle{5.5}}
\put(56.25,79.75){\color{blue}\circle{1.5}}
\put(38.75,51.25){\color{blue}\circle{1.5}}
\put(38.75,51.25){\color{blue}\circle{5.5}}
\put(74.75,109.75){\color{red}\circle{5.5}}
\put(74.75,109.75){\color{red}\circle{1.5}}
\put(74.75,109.75){\color{blue}\circle{9}}
\put(93.75,79.75){\color{red}\circle{5.5}}
\put(93.75,79.75){\color{red}\circle{1.5}}
\put(111.25,51.25){\color{red}\circle{5.5}}
\put(111.25,51.25){\color{red}\circle{1.5}}
\put(15,5){\makebox(0,0)[cc]{$\textsf{\textbf{A}}$}}
\put(56.25,5){\makebox(0,0)[cc]{{\color{green}$\textsf{\textbf{B}}$}}}
\put(92.5,5){\makebox(0,0)[cc]{{\color{green}$\textsf{\textbf{C}}$}}}
\put(138,5){\makebox(0,0)[cc]{$\textsf{\textbf{D}}$}}
\put(125,52.25){\makebox(0,0)[lc]{{\color{red}$\textsf{\textbf{E}}$}}}
\put(108,80.25){\makebox(0,0)[lc]{{\color{red}$\textsf{\textbf{F}}$}}}
\put(74.75,128){\makebox(0,0)[cc]{$\textsf{\textbf{G}}$}}
\put(42.5,80.25){\makebox(0,0)[rc]{{\color{blue}$\textsf{\textbf{H}}$}}}
\put(25.5,52.25){\makebox(0,0)[rc]{{\color{blue}$\textsf{\textbf{I}}$}}}
\end{picture}
&
$\quad \quad$&
\unitlength 0.30mm
\allinethickness{2pt}
\begin{picture}(103.67,92.00)
\put(0.00,80.33){{\color{red}\line(2,-1){40.00}}}
{{\color{red} \bezier{104}(40.00,60.33)(50.33,55.00)(50.00,40.33)}}
\put(50.00,40.33){{\color{red}\line(0,-1){40.00}}}
\put(102.00,80.33){{\color{green}\line(-2,-1){40.00}}}
{{\color{green} \bezier{104}(62.00,60.33)(51.67,55.00)(52.00,40.33)}}
\put(52.00,40.33){{\color{green}\line(0,-1){40.00}}}
\put(0.00,80.33){{\color{red}\circle{3.33}}}
\put(40.00,60.33){{\color{red}\circle{3.33}}}
\put(51.00,40.33){{\color{red}\circle{3.33}}}
\put(0.00,80.33){{\color{red}\circle{1}}}
\put(40.00,60.33){{\color{red}\circle{1}}}
\put(51.00,40.33){{\color{red}\circle{1}}}
\put(51.00,40.33){{\color{green}\circle{5}}}
\put(51.00,0.33){{\color{red}\circle{3.33}}}
\put(51.00,0.33){{\color{red}\circle{1}}}
\put(51.00,0.33){{\color{green}\circle{5}}}
\put(102.00,80.33){{\color{green}\circle{3.33}}}
\put(62.00,60.33){{\color{green}\circle{3.33}}}
\put(102.00,80.33){{\color{green}\circle{1}}}
\put(62.00,60.33){{\color{green}\circle{1}}}
\put(45.00,0.33){\makebox(0,0)[rc]{$\textsf{\textbf{A}}$}}
\put(45.00,40.33){\makebox(0,0)[rc]{$\textsf{\textbf{B}}$}}
\put(40.00,71){{\color{red}\makebox(0,0)[cc]{$\textsf{\textbf{C}}$}}}
\put(0.00,90){{\color{red}\makebox(0,0)[cc]{$\textsf{\textbf{D}}$}}}
\put(62.00,71){{\color{green}\makebox(0,0)[cc]{$\textsf{\textbf{E}}$}}}
\put(102.00,90.00){{\color{green}\makebox(0,0)[cc]{$\textsf{\textbf{F}}$}}}
\end{picture}
\\
$\;$\\
a)& b)&& c) \\
\end{tabular}
\end{center}
\caption{(Color online) Diagrammatical representation of interlinked contexts by Greechie (orthogonality) diagrams
(points stand for individual basis vectors, and entire contexts are drawn as smooth curves):
a) two tripods with a common leg;
b) three interconnected fourpods (this configuration with tripods would be
irrepresentable in three-dimensional vector space~\protect\cite{kalmbach-83,pulmannova-91});
b) two contexts in four dimensions interconnected by two link observables.
}
\label{2010-pc09-f1}
\end{figure}
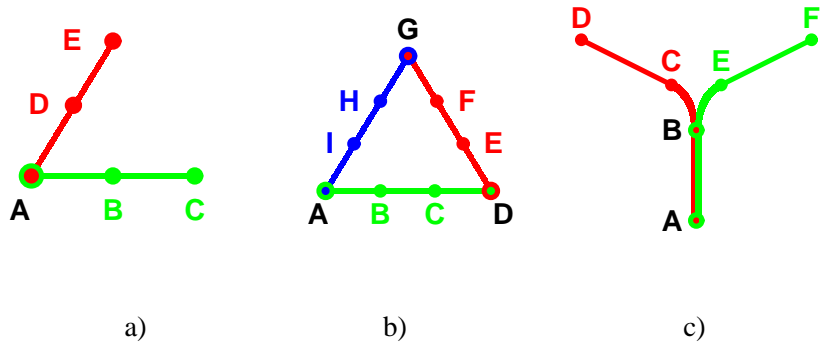

There exist ``explosion views'' of counterfactual configurations involving singlet or other correlated states of two
three- and more state particles which, due to the counterfactual uniqueness properties~\cite{svozil-2006-uniquenessprinciple},
are capable of indirectly testing the quantum contextuality
assumption~\cite{svozil:040102} by a simultaneous measurement of two complementary contexts~\cite{epr}.
For the sake of  explicit demonstration, consider Fig.~\ref{2010-pc09-f1} depicting
three orthogonality (Greechie) diagrams of such configurations of observables.
Every diagram is representable in three- or four-dimensional vector space.

For the configuration depicted in Fig.~\ref{2010-pc09-f1}a),
contextuality predicts that there exist experimental outcomes with
$
\textsf{\textbf{A}} \{  \textsf{\textbf{B}},\textsf{\textbf{C}} \}
\neq
\textsf{\textbf{A}} \{  \textsf{\textbf{D}},\textsf{\textbf{E}} \}
$.
As detailed quantum mechanical calculations~\cite{svozil:040102} show,  this is not predicted by quantum mechanics.

For the configuration depicted in Fig.~\ref{2010-pc09-f1}b),
contextuality predicts that there exist experimental outcomes with
$
\textsf{\textbf{A}} \{  \textsf{\textbf{B}},\textsf{\textbf{C}},\textsf{\textbf{D}} \}
\neq
\textsf{\textbf{A}} \{  \textsf{\textbf{G}},\textsf{\textbf{H}},\textsf{\textbf{I}} \}
$,
as well as
$
\textsf{\textbf{A}} \{  \textsf{\textbf{G}},\textsf{\textbf{H}},\textsf{\textbf{I}} \}
=
\textsf{\textbf{D}} \{  \textsf{\textbf{E}},\textsf{\textbf{F}},\textsf{\textbf{G}} \}
=1
$, and their cyclic permutations.

For the configuration depicted in Fig.~\ref{2010-pc09-f1}c),
contextuality predicts that there exist experimental outcomes with
$
\textsf{\textbf{A}} \{  \textsf{\textbf{B}},\textsf{\textbf{C}},\textsf{\textbf{D}} \}
\neq
\textsf{\textbf{A}} \{  \textsf{\textbf{B}},\textsf{\textbf{E}},\textsf{\textbf{F}} \}
$,
as well as
$
\textsf{\textbf{B}} \{  \textsf{\textbf{A}},\textsf{\textbf{C}},\textsf{\textbf{D}} \}
\neq
\textsf{\textbf{B}} \{  \textsf{\textbf{A}},\textsf{\textbf{E}},\textsf{\textbf{F}} \}
$.
Again, this is not predicted quantum mechanically~\cite{svozil:040102}.

Experiment will clarify and decide the contradiction
between the predictions by the contextuality assumption and quantum mechanics,
but it is not too unreasonable to suspect that the quantum predictions will prevail.
As a consequence, and subject to experimental falsification,
any {\it ad hoc} ``ontic'' contextuality assumption might turn out to be physically unfounded.

One may argue that quantum contextuality only ``appears'' if measurement configurations are encountered which do not allow a
set of two-valued states. The same might be said  for measurement configurations allowing only a ``meager'' set of two-valued states which
cannot be used for the construction of any homomorphic (i.e. preseving relations and operations among quantum propositions) embedding
into a classical (Boolean) algebra.
Alas, configurations of observables such as the one depicted  in Fig.~\ref{2010-pc09-f1}a)
are just subconfigurations of proofs of the Kochen-Specker theorem~\cite{kochen1}, in particular their
$\Gamma_2$ and $\Gamma_3$; so it would be difficult to imagine
why Fig.~\ref{2010-pc09-f1}a) feature context independence because of the experimenter takes into account only {\em two} contexts,
whereas context dependence is encountered when the experimenter has in mind, say, the {\em entire} structure of
all the 117 Kochen-Specker contexts contained in $\Gamma_3$.

\section{Context translation principle}

In view of the inapplicability of the quantum contextuality assumption and the fact that,
although quantized systems can only be prepared in a certain single context\footnote{
We would even go so far to speculate that
the ignorance of state preparation resulting in mixed states is an epistemic, not ontologic, one.
Thus all quantum states are ``ontologically'' pure.}
quantized systems yield measurement results when measured ``along'' different, nonmatching context,
one may speculate that the measurement apparatus must be capable of ``translating''
between the preparation context and the measurement context~\cite{svozil-2003-garda}.
Variation of the capabilities of the measurement apparatus to translate nonmatching quantum contexts
with its physical condition  yields possibilities to detect this mechanism.

In this scenario, stochasticity is introduced {\it via} the context translation process;
albeit not necessarily an irreversible,
irreducible one, as the unitary quantum state evolution (in-between measurements)
is  deterministic, reversible and one-to-one~\cite{everett}.
Nevertheless, one may further speculate
that, at least for finite experimental time series and for finite algorithmic tests,
any such quasi-deterministic form of stochasticity will result in very similar statistical behaviors
as is predicted for acausality.

Context translation might present an ``epistemic'' contextuality,
since the ``complete disposition  of the measurement apparatus'' (see Bell~\cite[Sec.~5]{bell-66})
may enter in the translation function $\tau$ formalizing the ``state reduction''
\begin{equation}
{\bm \rho}  \longrightarrow \tau_{\textsf{{D}}\left(\textsf{\textbf{X}},\textsf{\textbf{Y}}\right)}  ({\bm \rho}) \in S_\textsf{\textbf{X}},
\label{2010-pc09-e1}
\end{equation}
where
${\bm \rho}$ stands for the quantum state,
$S_\textsf{\textbf{X}}$ for the spectrum of the operator $\textsf{\textbf{X}}$,
$\textsf{{D}}$ for the ``disposition of the apparatus,''
$\textsf{\textbf{X}}$ for the observable
and $\textsf{\textbf{Y}}$ for the context.

In general, even in the absence of some concrete ``translation mechanism,''
$\tau$ is subject to some probabilistic constraints, such as Malus' law~\cite{zeil-bruk-99}.
In order to be able to account for the nonlocal quantum correlation functions even at space-like separations~\cite{wjswz-98}
$\tau$ should also be nonlocal.
Ideally, if preparation and measurement context match, and if
${\bm \rho}$ is in some eigenstate $\textsf{\textbf{E}}_i$ of $\textsf{\textbf{X}}$ with an associated eigenvalue $x_i$,
then  Eq.~(\ref{2010-pc09-e1}) reduces to its context and apparatus independent form
$\textsf{\textbf{E}}_i  \longrightarrow \tau_{\textsf{{D}}\left(\textsf{\textbf{X}},\textsf{\textbf{Y}}\right)}
(\textsf{\textbf{E}}_i) =  \tau_{\textsf{\textbf{X}}}  (\textsf{\textbf{E}}_i) =x_i$ for all $\textsf{\textbf{E}}_i$
in the spectral sum
$\textsf{\textbf{X}} = \sum_i x_i \textsf{\textbf{E}}_i$.
This reduction postulate appears to be the reason for an absence of contextuality in the ``explosion view'' type configurations
discussed above.

For all the other cases, the measurement apparatus will introduce a stochastic element
which, in this scenario, is the reason for the quantum indeterminism of individual events.
Of course, the {\em degree of stochasticity} will depend on the context mismatch,
and on the ``disposition of the apparatus.''
But again, as for the {\em ad hoc} ``ontic'' type of contextuality discussed above,
in no way can the measurement outcome of an individual particle be completely determined
by a pre-existing element of physical reality~\cite{epr} of that particle alone.
In this sense, as only observables associated with one context have a definite value and all other observables have none,
one is  lead to a quasi-classical ``effective value indefiniteness,''
giving rise to a natural classical theory not requiring value definiteness.

\section{Summary}

We have discussed the ``current state of affairs'' with regard to the interpretation of quantum value indefiniteness,
and the limited  operationalizability of its interpretation in terms of {\em context dependence (contextuality)} of observables.
Of course, due to complementarity, quantum counterfactuals are not directly simultaneously measurable;
and thus --- despite the prevalence of counterfactuals in quantum information, communication and computation theory ---
anyone considering their physical existence is,
to paraphrase von Neumann's words~\cite{von-neumann1}, at least empirically, {\em ``in a state of  sin.''}

In any case, the absence of classical interpretations of the quantum formalism,
and in particular the strongest expression of it
---
the absence of any global truth function for quantum systems of three or more mutually exclusive outcomes
---
presents the possibility to render a quantum random number generator
by preparing a quantum state in a particular context and measuring it in another.
As has been pointed out already by Calude and the author~\cite{2008-cal-svo}, the resulting measurement outcomes
are ``quantum certified'' (i.e., true with respect to the validity of quantum mechanics) and do not correspond to any
pre-existing physical observable of the ``isolated'' individual system before the measurement process.
Exactly how this kind of quantum oracle for randomness operates remains open.
One may hold that, somehow, due to the lack of determinacy, this type of randomness emerges ``out of nowhere'' and essentially is
irreducible~\cite{zeil-99,zeil-05_nature_ofQuantum}.
One may also put forward the idea that, at least when complementarity is involved, quantum randomness is rendered by
a quasi-classical context translation which maps an incompatible preparation context into some outcome,
thereby introducing stochasticity.
In any case, for all practical purposes, the resulting oracles for randomness, when subjected to tests~\cite{PhysRevA.82.022102},
might be ``hardly differentiable'' from each other even asymptotically.

\section*{Acknowledgements}
The author would like to thank two anonymous Referees for very valuable observations and suggestions.


%

\end{document}